\documentclass{aa}
\usepackage{graphicx}
\usepackage{txfonts}
\usepackage{color}
\usepackage{multirow}
\begin{document}

\title{Precise Wavefront Correction with an Unbalanced Nulling Interferometer for Exo-Planet Imaging Coronagraphs}

   \author{J. Nishikawa
          \inst{1,2},
          L. Abe
          \inst{3,2}
				\thanks{current address: Laboratoire Hippolyte Fizeau, UMR 6525 Universite de Nice-Sophia Antipolis, 
                             Parc Valrose, F-06108 Nice, France, \email{Lyu.Abe@unice.fr}}
          N. Murakami
          \inst{1},
          \and
          T. Kotani\inst{4}
				\thanks{current address: LESIA, Observatoire de Paris, section Meudon, 5 Place Jules Janssen,
                 92195 Meudon, France, \email{takayuki.kotani@obspm.fr}}
          }

   \offprints{J. Nishikawa}

   \institute{MIRA project, National Astronomical Observatory of Japan,
              Mitaka, Tokyo 181-8588, Japan\\
              \email{jun.nishikawa@nao.ac.jp}
              \email{naoshi.murakami@nao.ac.jp}
             \and
              Extrasolar Planet Project Office, National Astronomical Observatory of Japan,
              Mitaka, Tokyo 181-8588, Japan
             \and
              Division of Optical and Infrared Astronomy, National Astronomical Observatory of Japan,
               Mitaka, Tokyo 181-8588, Japan 
             \and
             Max-Planck-Institut f\"{u}r Radioastronomie, Auf dem H\"{u}gel 69, D-53121 Bonn, Germany\\
             }

\date{Received 2006 May 22; accepted 2008}

\abstract
{Very high dynamical range coronagraphs targeting direct exo-planet detection \rm ($\rm{10^{9}\sim10^{10}}$ \rm contrast) at small
angular separation (few $\lambda/D$ units) usually require an input wavefront quality on the order of ten thousandths
of wavelength RMS.}
{We propose a novel method based on a pre-optics setup that behaves partly as a low-efficiency coronagraph, and partly
as a high-sensitivity wavefront \rm aberration \rm compensator (phase and amplitude). The combination of the two effects 
results in a highly accurate corrected wavefront.}
{First, an (intensity-) unbalanced nulling interferometer (UNI) performs a rejection of part of the wavefront electric
field. Then the recombined output wavefront has its input \rm aberrations \rm magnified. Because of the unbalanced
recombination scheme, \rm aberrations \rm can be free of phase singular points (zeros) and can therefore be compensated by a
downstream phase and amplitude correction (PAC) adaptive optics system, using two deformable mirrors.}
{In the image plane, the central star's peak intensity and the noise level of its speckled halo are reduced by the
UNI-PAC combination: the output-corrected wavefront \rm aberrations \rm can be interpreted as an improved compensation of the
initial (eventually already corrected) incident wavefront \rm aberrations. \rm }
{The important conclusion is that \rm not all \rm the elements in the optical setup using UNI-PAC \rm need \rm to reach the
$\lambda/10000$ rms surface error quality.}


\authorrunning{Nishikawa et al.}
\titlerunning{Wavefront Correction Nulling Coronagraph}
 \maketitle

\section{Introduction}

Optical coronagraphy in space is one of the most useful methods to achieve high dynamic range observations for the
direct detection of extra-solar planets (e.g., \cite{Coul05}; \cite{Aime06}). A coronagraph can reduce a central star
intensity and its diffracted halo light around the star where exo-planets would appear. Several coronagraph designs
using  advanced focal plane masks (\cite{Guyo99}; \cite{Kuch02}; \cite{Baba02}; \cite{Riau03}) have been reported which
reduce the light energy, sometimes to zero in theory, inside a re-imaged pupil plane called a Lyot stop, and also at a
final image. Some \rm techniques \rm of nulling interferometry, which has been mainly considered for mid-infrared
long-baseline interferometers (\cite{Menn05}), \rm has been applied \rm to optical coronagraph in a single telescope using
rotation-shearing interferometers (\cite{Baud00}; \cite{Tavr05}) or lateral-shearing interferometers with overlapped or
separated sub-apertures (\cite{Shao04}; \cite{Nish05}). Pupil function modification can reduce only the halo intensity
by using shaped or apodized pupils (\cite{Kasd05}; \cite{Gali05}). Wavefront phase control by an adaptive optics (AO)
system can reduce the halo intensity of a limited area, called ``dark hole'' (\cite{malbet95}). Multiple-stage coronagraphs
(\cite{Aime04}; \cite{Toll05}), and combinations of these methods, or pre-optics schemes have been proposed to achieve
a higher dynamic range (\cite{Nish05}; \cite{NishM06}; \cite{Abe06}).

In any case, the required dynamic range of ${\rm 10^{9}\sim10^{10}}$ \rm for direct detection of earth-like planets by optical
coronagraphs can be achieved only with a very high quality wavefront of $\lambda$/10000 rms and an intensity uniformity
of 1/1000 rms (\cite{Kuch02}; \cite{Lowman}). Indeed, wavefront \rm aberrations \rm throughout the optical train produce a
so-called speckle halo noise in the image plane (e.g., \cite{Bord06}) that prevents direct detection of the planet if
these speckles are too bright. Manufacturing or polishing accurate mirrors would be a good direction for small mirrors,
while it would not be easy for a telescope primary to reach the required surface accuracy. Then an AO system would be
used to achieve the required wavefront accuracy where both measurements and corrections of the wavefront are important.
Giant or young planets brighter than $10^{6}$ contrast are the target of ground-based telescopes in the infrared
wavelength where wavefront correction by an AO system is also a key technology \rm for use against \rm atmospheric turbulence (e.g.,
\cite{Maki06}; \cite{Sera06}; \cite{Tamu06}).

In an AO system, a deformable mirror (DM) is controlled by a signal from a wavefront sensor (WFS). Recently, control
accuracy of a DM has been shown at a level of $\lambda$/10000 rms (e.g., \cite{Brow03}; \cite{Evan05}), while some
commercial wavefront sensors have been reported to achieve absolute wavefront measurement accuracy of $\lambda$/1000
rms by conventional Shack-Hartmann method (e.g., HASO HP 26 from Imagine Optic). The same goes for interferometric
sensing, where repeatability on the order of $\lambda$/10000\,rms is commonly achieved (e.g., Zygo VeriFire AT). In the
high dynamic range coronagraph regime, however, a non-common path error problem, i.e., differences between optics of
the WFS and the main path for the star and the planet images, should be considered in the pupil plane WFS such as the
Shack-Hartmann sensor, the interferometer, or a curvature sensor (e.g., \cite{Guyo06}).

Solutions for the non-common path error problem are focal plane WFSs (\cite{Codo04}; \cite{Bord06}; \cite{Give06}) or
speckle nulling, \rm which features \rm focal plane speckle measurement and iterative wavefront control. A dynamic range of
$10^{9}$ within the dark hole has been achieved after a few \rm thousand iterations \rm by an AO system (\cite{Bala06}) 
where the wavefront correction \rm approaches \rm the requirement.

These focal-plane sensing approaches, however, meet other problems. When \rm the speckle intensity level
gets closer to the planet intensity, long exposure times are required to obtain \rm better S/N measurements of the residual
wavefront \rm aberrations. In addition, \rm the wavefront stability of the whole telescope system must be guaranteed during \rm 
this time lapse. \rm A multiple stage coronagraph reducing speckles with AOs at every stage (\cite{Toll05}) faces the same
problem. In the case of multiple stages \rm with approaches \rm using nulling coronagraphs or nulling interferometers, no wavefront
sensing has been considered before (e.g. \cite{Nish05}; \cite{NishM06}). But the same wavefront sensing problems would
exist, i.e. the intensity in the pupil plane (at the Lyot stop) can be zero or very low, exhibiting a random \rm intensity
pattern \rm where phase singularity at zero intensity points makes wavefront measurements and corrections difficult. Thus
in the very high dynamic range optics for direct detection of exo-planets, maintaining wavefront quality remains an
issue as well as the development of coronagraphs.

\begin{figure}
\centering\includegraphics[width=\columnwidth]{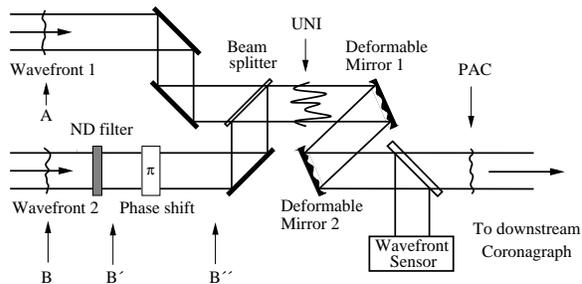} \caption{Schematic of a possible optical layout for
the UNI-PAC method (unbalanced nulling interferometer followed by a phase and amplitude correction
adaptive optics). The various optical planes are those referred to in the text.} \label{fig_optics}
\end{figure}

We propose a pre-optics scheme for coronagraphs, using a combination of an unbalanced nulling interferometer (UNI) and
a two-DM AO system (Fig.\ref{fig_optics}) for Phase and Amplitude Correction (PAC). Similarities can be found with
previous studies, i.e. pre-optics by a balanced nulling interferometer (\cite{Nish05}; \cite{NishM06}), coronagraphic
pre-optics with non-zero amplitude distribution in the pupil after an interference and complex amplitude correction
(\cite{Abe06}), and a multiple-stage wavefront correction by commercial quality AO systems (\cite{Toll05}).

The pre-optics concept using a balanced nulling interferometer in front of other coronagraphs was shown to be effective
in absorbing part of the required dynamic range (\cite{Nish05}; \cite{NishM06}) but could not maintain
sufficiently high wavefront \rm aberrations \rm after the interference to be corrected again by a secondary AO system. As a
consequence, it puts some very drastic constraints on the precision and accuracy of the upstream AO system. Abe et al.
(2006) showed that a pre-optics, that uses a pseudo-coronagraphic stage and a complex amplitude filter at the Lyot
plane, can produce a flat and almost uniform wavefront, in spite of central obscuration and spider-arm patterns (i.e.
large amplitude \rm aberrations \rm at the entrance pupil). We extended this concept in order to facilitate the phase and
amplitude measurement/correction after a nulling stage, which is usually the bottleneck of high contrast coronagraphs.
In the proposed concept, the UNI aims at simultaneously performing the absorption of part of the dynamic range, and
the magnification of wavefront \rm aberrations \rm without phase singularities for later compensation with an AO system (i.e.
the PAC), so that the star's speckled halo is effectively attenuated.

\rm This paper introduces the principle of the UNI-PAC concept, \rm  a novel method that \rm avoids the non-common path error
problem and the low intensity or phase singularity problem. \rm The overview, formalization, and simulation of the 
concept \rm are \rm addressed in Sect.\,\ref{sect:overview}, Sect.\,\ref{sect:formalism}, and Sect.\,\ref{sect:simulation},
respectively, followed by a discussion in Sect.\,\ref{sect:discussion}. Conclusions \rm are \rm drawn in
Sect.\,\ref{sect:conclusion}.

\section{Overview of the concept}
\label{sect:overview}

\begin{figure}
\centering \includegraphics[width=0.9\columnwidth]{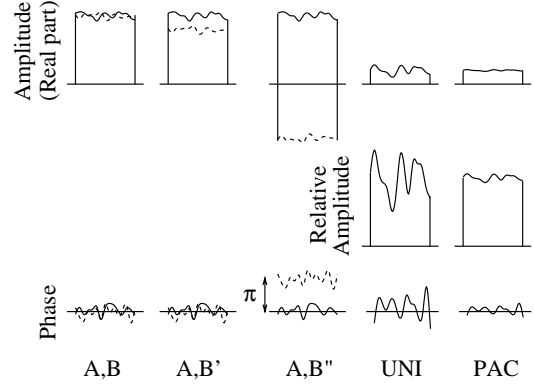} 
\caption{\rm Schematic of wavefront phase and
amplitude magnification in the process of  the UNI-PAC method,
  where only the real part is indicated for the amplitude.
    In planes A, B, B', B'' solid and dashed lines show wavefront 1 and 2, respectively. In planes UNI and PAC
    the combined wavefront is shown. \rm }
\label{fig_amp_phase}
\end{figure}

\rm A \rm rough illustration of the UNI-PAC process \rm is as follows. Figure\,\ref{fig_optics} shows a
possible optical concept of the UNI-PAC system, and Fig.\,\ref{fig_amp_phase} depicts the associated effects on the
amplitude and phase in the pupil plane at different points in the set-up. The UNI uses two wavefronts extracted from
a collimated beam of a telescope. This beam can be compensated by an AO system, for instance, at a reference level of
$\lambda/1000$ rms that we will use throughout this paper.

In our optical scheme, the UNI stage is intended to produce two effects. Firstly, it reduces the intensity of the 
star image by, e.g., an order of 1/100, but has no effect on the speckle halo level. This corresponds to an 
amplitude reduction of the unaberrated electric field by a partial nulling in the pupil plane.

A second effect of the UNI is the wavefront aberration magnification (for instance, from $\lambda/1000$ to 
$\lambda/100$), which is in fact a consequence of the partial nulling, and one of the most important 
phenomenon in the UNI-PAC concept. It is consistent with that the difference between the attenuated star 
intensity and the speckle level becomes smaller.

The next stage is the PAC, that is, a second AO system whose goal is to compensate for the magnified phase and
amplitude aberrations (from $\lambda/100$ to $\lambda/1000$ in our example). The corresponding effect in the
image plane is the reduction of the speckled halo level. The PAC stage can effectively control the wavefront because
the UNI stage has maintained an almost uniform intensity level in the pupil plane (i.e. with no phase singularities).

In our example, before the UNI stage, the initial intensity ratios of the star's peak intensity, the speckle halo,
and the planet are ${1:10^{-7}:10^{-9}}$ under a wavefront aberration of about $\lambda/1000$\,rms. After the UNI,
the ratios become ${10^{-2}:10^{-7}:10^{-9}}$. Here the starlight is reduced but the speckle is still there. Then, after
the PAC stage they become ${10^{-2}: 10^{-9}: 10^{-9}}$. Thus the UNI-PAC has reduced both the star and its
speckled halo intensity levels by a factor of 100.

A downstream coronagraph can then be used to further decrease the star intensity to detect the planet.
Note that the coronagraph does not require $\lambda/10000$ optics quality any more, but only about
$\lambda/1000$ which is enough to perform a ${10^{-7}}$ starlight attenuation, and keep the speckle 
level down to the planet level.

We also recall that the aberration level was $\lambda/1000$ both at the UNI-PAC and the first AO 
system locations (which can also include the non-common path errors of the AO systems).
Despite an overall optical quality of $\lambda/1000$, we have reached a dynamic range of $10^9$,
meaning that the wavefront quality is "virtually" improved beyond the AO system capabilities.
A big advantage of using the UNI-PAC is this specification relaxation for the entire optics of the coronagraph,
which is probably key to achieving the planet detection system in a cost-effective way.
\rm

\begin{figure}
    \centering\includegraphics[width=0.7 \columnwidth]{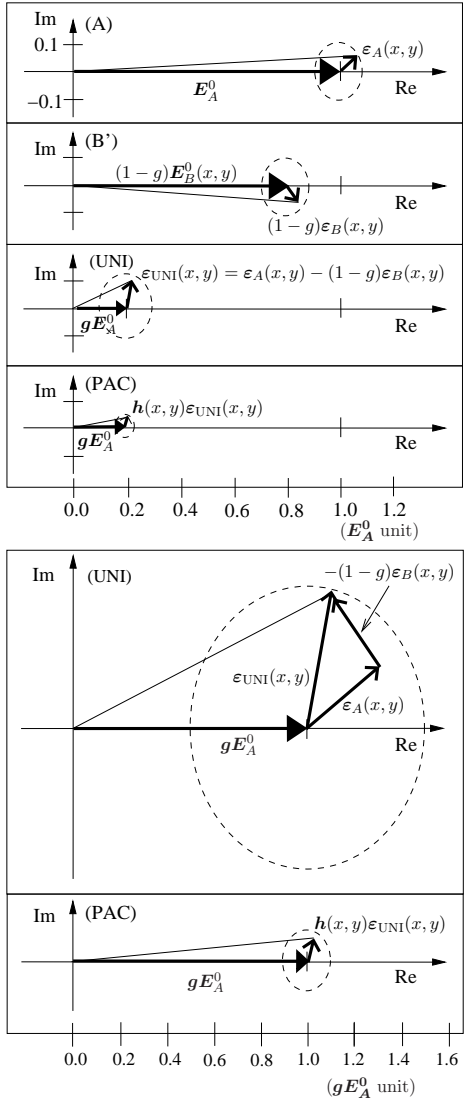}
\caption{Complex amplitude \rm at \rm a point $(x,y)$ in the pupil plane under the processes of the UNI-PAC method. 
In each panel, an arrow from the origin indicates the unaberrated amplitude, $E^0$, and the
complex aberration component, $\varepsilon$, is shown by other small arrow(s), where 
the distribution area of the complex amplitude is drawn by a dashed-lined ellipse. 
Traditional phase aberration, $\theta$,
is indicated by an arc. The upper four panels indicate the complex amplitude of beam A, beam B with ND filter
before $\pi$ phase shift, UNI, and finally PAC planes respectively from top to bottom, in $E^0_\mathrm{A}$ units. The
two lower panels are similar to the last two frames of the upper panel, except that the vectors have been normalized by
the factor $gE^0_\mathrm{A}$, where $g$=0.2. The normalized aberration is magnified by $1/g$ by the UNI (UNI output
plane), and compensated by a factor $h(x,y)$ at the PAC output plane.}
    \label{fig_complex}
\end{figure}

\section{Formalism}  \label{sect:formalism}
\rm
\subsection{Definitions and preparations}  \label{sect:definition}

\rm We \rm consider the complex amplitude of an electric field for each point on a pupil plane whose coordinate is $(x,y)$
\rm defined by the sum of an unaberrated constant amplitude $E^{0}$ and a complex aberration term $\varepsilon$ gathering
both amplitude and phase aberrations,
\begin{equation}
  E(x,y) = E ^{0} + \varepsilon (x,y) \, .
\label{Eq_E}
\end{equation}
We prefer to use this expression rather than the somewhat classical ${E=E^0\,(1+\alpha)\, \rm{e}^{\rm{i}\theta}}$
notation (where $\alpha $ and $\theta $ are the amplitude and the phase aberrations, respectively).
Indeed, the ratio $\varepsilon(x,y)/E^{0}$, ``the normalized complex aberration'', is a useful indicator of \rm the
wavefront quality \rm in our algebraic derivations. Also, under small wavefront aberration conditions, its
real and imaginary parts are almost equal to $\alpha (x,y) $ and $\theta (x,y) $, respectively (the correspondence
between these notations can be found in Appendix~\ref{Appendix_A}). That way, we can treat ``the
normalized complex aberration'' as the wavefront aberration. We can assume $E^{0}$ is a real number and under a very
small wavefront aberration condition
\begin{eqnarray} \left\{ \begin{array}{lll}
    ~~~~~~E^{0}  &=& \big\langle E(x,y) \big\rangle  \\[2mm]
    \big\langle \varepsilon (x,y) \big\rangle &=& 0 \, ,
  \label{Eq_E0}
\end{array} \right. \end{eqnarray}
where \rm $\big\langle \ \big\rangle$ is a spatial ensemble average within the pupil plane.

We use notations for the intensity of each complex amplitude of electric field as 
\begin{eqnarray} \left\{ \begin{array}{lll}
    I^{0} &=& | \, E^{0} \, |^2 \\[1mm]
    I^{\varepsilon}(x,y) &=& | \, \varepsilon(x,y) \, |^2  \, , 
  \label{Eq_I0}
\end{array} \right.  \end{eqnarray}
\rm where \rm $|~~|$ is the absolute value for real numbers or the modulus for complex numbers.
\rm $I^{\varepsilon}$ can be observed as a residual intensity in the pupil plane when the unaberrated
component is sufficiently attenuated. \rm The average \rm of the residual intensity \rm is proportional to the average speckle
\rm background \rm intensity in the image plane.

\rm In the following equations, \rm the notations described above will \rm include \rm a suffix, indicating the
location where the wavefront is evaluated, that is, \rm one of six planes (\rm A, B, \rm B', B'', UNI and then PAC
planes) in the optical layout of the concept of Fig.~\ref{fig_optics}. \rm In Fig.\,\ref{fig_complex} a vector
representation of the complex amplitude is shown for each plane in the UNI-PAC. \rm

We consider two wavefronts $E _\mathrm{A}(x,y) $ and $E _\mathrm{B}(x,y) $ at the input of our system generated from a
collimated beam of a single telescope. These wavefronts are corrected in advance by an AO system before the UNI stage,
but \rm aberrations (including non-common path errors) \rm still remain written, 
\begin{eqnarray} \left\{ \begin{array}{l}
  E_\mathrm{A} (x,y) =  E _\mathrm{A}^0 + \varepsilon _\mathrm{A} (x,y) ~ \\[2mm]
  E_\mathrm{B} (x,y) =  E _\mathrm{B}^0 + \varepsilon _\mathrm{B} (x,y) \, .
  \label{Eq_EAEB}
\end{array} \right. \end{eqnarray}
\rm Any \rm AO system has a limited number of actuators, hence usually limiting the correction to lower spatial frequency ranges.
This is a common situation of AO systems, but fortunately the low spatial frequency region corresponds to the relevant
inner area in the image plane where the exo-planets would be observed. In this paper, we assume that a 
\rm rather ordinary quality of $\lambda/1000$\,rms level for \rm
the AO system is used on the telescope side before our concept, and another \rm AO of \rm the same quality is again
adopted in the latter \rm half \rm of our concept \rm (the PAC stage). \rm We also consider how the combination of the unbalanced
nulling interferometer and the AO system works for this limited inner region.

\rm For the two wavefronts, \rm we assume \rm the same unaberrated amplitude, 
\begin{eqnarray} \left\{ \begin{array}{lll}
  E_\mathrm{A}^0 &=& E_\mathrm{B}^0 \\[2mm]
  I_\mathrm{A}^0 &=& I_\mathrm{B}^0 \, ,
  \label{Eq_EA0}
\end{array} \right. \end{eqnarray}
and complex aberrations of \rm similar level \rm (variance) \rm as 
\begin{equation}
  \big\langle I_\mathrm{A}^{\varepsilon}(x,y) \big\rangle = \big\langle I_\mathrm{B}^{\varepsilon}(x,y) \big\rangle \, .
  \label{Eq_eA}
\end{equation}

\subsection{Unbalanced nulling interferometer (UNI) process }
\rm The first effect of the UNI stage is a starlight attenuation by a nulling interference of the two beams.
For the wavefront $E _\mathrm{B}$ (plane B) \rm the modulus of the complex amplitude is slightly reduced,
for \rm instance \rm by an ND filter. At the output of this ND filter, the complex amplitude becomes, 
\begin{equation}
  E_\mathrm{B'} (x,y)  =  (1-g)~ \bigl( E _\mathrm{B}^0 + \varepsilon _\mathrm{B} (x,y) \bigl)
  \label{Eq_EB'}
\end{equation}
where $g$ is a small fraction \rm of unity, \rm i.e. the unbalance factor. The beam is \rm then \rm phase-shifted, such that 
\begin{eqnarray}
  E_\mathrm{B''} (x,y) &=& e^{\mathrm{i}\phi} (1-g)~ \bigl( E _\mathrm{B}^0 + \varepsilon _\mathrm{B} (x,y) \bigl)  \nonumber \\
  &=&-\, (1-g)~ \bigl( E _\mathrm{B}^0 + \varepsilon _\mathrm{B} (x,y) \bigl)  \, .
  \label{Eq_EB''}
\end{eqnarray}
\rm where we considered $\phi=\pi$. \rm 

The two wavefronts \rm $E_\mathrm{A}(x,y)$ and $E_\mathrm{B''}(x,y)$ \rm are summed for destructive interference, so that at
the UNI output plane (subscripted UNI), \rm and using Eq.(\ref{Eq_EA0}), \rm 
\begin{eqnarray}
   E _{\mathrm{UNI}}(x,y)  &=&  g \, E_\mathrm{A}^0 + \varepsilon _\mathrm{A}(x,y) - (1-g) ~ \varepsilon _\mathrm{B}(x,y) \, ,
\label{Eq:EUNI}
\end{eqnarray}
where we \rm omitted \rm the reflection and transmission ratio of the beam combiner.

\rm There is one point we need to clarify here: the important effect we want to achieve is to obtain an 
appropriate value for the modulus of the unaberrated component of the residual complex amplitude 
after the beam combination. In other words, it does not matter whether this residual is purely real or not. 
Therefore, in Eq.\,(\ref{Eq_EB''}), the factor ${e^{\mathrm{i}\phi} (1-g)}$ can be complex, as long 
as we can control either $g$ or $\phi$ (or both) to produce the desired unaberrated modulus amplitude for the 
combined beam. In the following, we assume ${\phi=\pi}$ for simplicity. \rm

\rm We consider that defects of the optics for the unbalancing as well as for the phase shift optics are 
negligible, or included in $\varepsilon_\mathrm{B} $. The aberrations of the beam combiner and the 
two optical paths in the UNI are treated in the 
same way using $\varepsilon_\mathrm{A} $ and $\varepsilon_\mathrm{B} $. 

The unaberrated \rm complex amplitude \rm of Eq.\,(\ref{Eq:EUNI}) and its intensity \rm become 
\begin{equation} \left\{ \begin{array}{lll}
  E_{\mathrm{UNI}}^0  &=& g \, E_\mathrm{A}^0 \\[2mm]
  I^0_{\mathrm{UNI}} &=& g^2 \, I^0_\mathrm{A} \, .
    \label{Eq_E0UNI}
\end{array} \right.
\end{equation}
At this point, the starlight intensity is reduced by $g^2$, e.g., if ${g=0.1}$ the starlight
is attenuated by 1/100. \rm The complex aberration term \rm is almost the sum of the two wavefronts \rm as 
\begin{equation} \left\{ \begin{array}{lll}
    \varepsilon _{\mathrm{UNI}}(x,y) &=& \varepsilon _\mathrm{A}(x,y) - (1-g) ~ \varepsilon _\mathrm{B}(x,y)    \\[2mm]
    \big\langle I^\varepsilon_{\mathrm{UNI}}(x,y) \big\rangle
                                               &=&   \big(2-2g+g^2\big) ~ \big\langle I^\varepsilon_\mathrm{A}(x,y) \big\rangle \, ,
\label{Eq_eUNI}  
\end{array} \right. \end{equation}
\rm where Eq.(\ref{Eq_eA}) and the independence of $\varepsilon _\mathrm{A}(x,y)$ and $\varepsilon_\mathrm{B}(x,y)$
are used for the latter equation. This situation corresponds to a ``worst case" because if the aberrations are perfectly
correlated so that ${\varepsilon _\mathrm{A}(x,y)=\varepsilon_\mathrm{B}(x,y)} $, then 
${I^\varepsilon _\mathrm{UNI}(x,y)=g^2\,I^\varepsilon_\mathrm{A}(x,y)} $, meaning that the aberration term is well canceled as 
well as the unaberrated amplitude. Therefore,  we may actually get a better result in practice than the Eq.\,(\ref{Eq_eUNI}) 
if the wavefronts partially correlate (see Appendix~\ref{Appendix_B}). We continue with the worst case formula 
proving the UNI-PAC process works well. 

In order to evaluate the dynamic range improvement by the UNI stage, the unaberrated and the aberrated 
component of the star should be compared with the planet. We introduce the intensity of the planet $I^p_A $, 
where we only consider the unaberrated component, since its aberration component intensity is very small.
We adopt the following expression for the planet intensity after the UNI as
\begin{eqnarray}
     I^{p}_\mathrm{UNI} = (2-2g+g^2)~ I ^{p} _\mathrm{A} \,
  \label{IpUNI}
\end{eqnarray}
by averaging for the constructive and the destructive conditions of the interferometer (see Appendix C).
Then using Eqs.\,(\ref{Eq_E0UNI}), (\ref{Eq_eUNI}), and (\ref{IpUNI}), the dynamic range improvement through the UNI
stage is expressed as
\begin{equation} \left\{ \begin{array}{lll}
  ~~~\frac{\textstyle ~I^0_\mathrm{UNI}~}{\textstyle I^{p}_\mathrm{UNI}}
      &=& \frac{\textstyle g^2}{\textstyle ~2-2g+g^2~} \frac{\textstyle ~I^0_\mathrm{A}~}{\textstyle I^{p}_\mathrm{A}} \\[4mm]
  \frac{\textstyle \big\langle I^\varepsilon_\mathrm{UNI}(x,y)\big\rangle}{\textstyle I^{p}_\mathrm{UNI}}
      &=&  \frac{\textstyle \big\langle I^\varepsilon_\mathrm{A}(x,y)\big\rangle}{\textstyle I^{p}_\mathrm{A}} \, ,
  \label{Eq_I0UNI/IpUNI}  
\end{array} \right. \end{equation}
where the latter equation shows that the averaged aberration intensity is not affected by the UNI, while
the star intensity gets closer to the planet intensity by a factor of $g^2/(2-2g+g^2) $.

\begin{figure}
    \centering
    \includegraphics[width=0.85 \columnwidth]{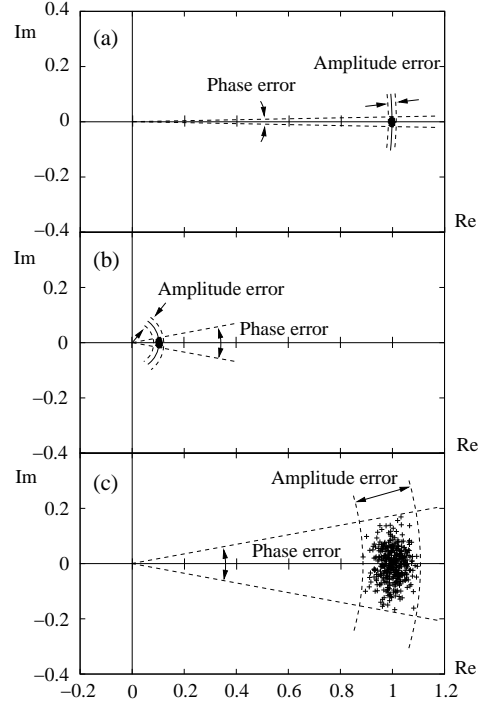}
\caption{Wavefront aberration magnification by the unbalanced nulling interference. (a) Complex amplitude of 400 points
is simulated in a wavefront whose phase aberration rms is 0.0063 radian and amplitude aberration rms is 0.004. (b)
After the unbalanced nulling interference with a no-aberration wavefront with an amplitude ratio of 0.9, the phase
aberration is easily seen as magnified, and (c) the amplitude aberration magnification is also obvious when the complex
numbers are normalized by the unaberrated wavefront electric field, which is 0.1$E^0$ in this case if we represent the
original unaberrated field as $E^0$. The magnified aberrations seen in panel (c) can be corrected again to the same
quality as (a) by an AO system in the next stage. } \label{fig_error_magnification}
\end{figure}

\subsection{Wavefront aberration magnification by UNI stage}
The important phenomenon at the UNI stage we must emphasize is the wavefront aberration magnification. \rm
The normalized complex \rm aberration \rm after the UNI \rm stage can be evaluated,
using Eqs.~(\ref{Eq_E0UNI}) and (\ref{Eq_eUNI}),
\begin{equation} \left\{ \begin{array}{lll}
   ~~~\frac{\textstyle \varepsilon _{\mathrm{UNI}}(x,y)} {\textstyle E_{\mathrm{UNI}}^0}
    &=&  \frac{\textstyle ~1~}{\textstyle g} ~\frac{\textstyle  \varepsilon _\mathrm{A}(x,y) 
             - (1-g) ~ \varepsilon _\mathrm{B}(x,y) } {\textstyle E _\mathrm{A}^0} \\[3mm]
   \bigg\langle \bigg| \frac{\textstyle \varepsilon _{\mathrm{UNI}}(x,y)} {\textstyle E_{\mathrm{UNI}}^0} \bigg|^2 \bigg\rangle
    &=&  \frac{\textstyle 2-2g+g^2} {\textstyle g^2} 
        ~\bigg\langle \bigg|\frac{\textstyle  \varepsilon _\mathrm{A}(x,y)} {\textstyle E _\mathrm{A}^0} \bigg|^2 \bigg\rangle \, .
   \label{Eq_eUNI/E0UNI}
\end{array} \right. \end{equation}
Here the UNI magnifies the normalized complex aberration by a factor $1/g$ (recalling that $g$\,$<$\,1) compared to the
initial wavefront aberrations, or, in average by $\sqrt{2-2g+g^2}{/g}$.

In order to obtain an appropriate magnification of the wavefront aberration,
we chose a sufficiently large value of the unbalance factor $g$, \rm
so that $E_{\mathrm{UNI}}(x,y)$ always has a positive intensity (modulus) \rm without any phase singularities,
 ${|E_{\mathrm{UNI}}(x,y)| > 0} $. \rm A little tighter constraint \rm is also suitable, 
\begin{equation}
    | \, E _\mathrm{UNI}^0 \, | \, > \, | \, \varepsilon _\mathrm{UNI}(x,y) \, | \, , 
    \label{Eq_non-zero_condition2}
\end{equation}
which means \rm $E_{\mathrm{UNI}}(x,y)$ is distributed \rm over \rm half of the complex plane excluding the origin \rm (see Fig.\,\ref{fig_complex}).
Within the condition of this equation, the smaller $g$, the more the star signal is attenuated and the aberration is magnified.

An example of the aberration magnification by the UNI stage is shown in Fig.\ref{fig_error_magnification}, 
where one wavefront has $\lambda/1000 $ phase, 0.004 amplitude aberrations in rms, and an unbalance factor ${g=0.1}$. 
The initial aberration is ${|\varepsilon _\mathrm{A}(x,y)/E^0_\mathrm{A}|<0.02} $ and the magnified one is
${|\varepsilon _\mathrm{UNI}(x,y)/E^0_\mathrm{UNI}| < 0.2}$
which will be compensated again to less than 0.02 by the PAC stage.
Although ${g=0.02} $ can be adopted just to satisfy Eq.(\ref{Eq_non-zero_condition2}), a magnified wavefront
aberration close to unity ${|\varepsilon _\mathrm{UNI}(x,y)/E^0_\mathrm{UNI}| \sim 1}$ would be less favorable
to perform the wavefront correction at the PAC stage.

If ${g \rightarrow 0}$, the unaberrated amplitude tends to completely cancel out,
${E _{\mathrm{UNI}}^0 \rightarrow 0}$, and the nulling result consists of only the aberration terms,
${E_{\mathrm{UNI}}(x,y)=\varepsilon _\mathrm{A}(x,y) - \varepsilon _\mathrm{B}(x,y)}$ which are distributed
around the origin in the complex plane.
\rm In this case, the residual \rm intensity in the pupil plane takes the form of a speckled field with a \rm small \rm inhomogeneous
\rm amplitude \rm and \rm random phase \rm fluctuations, resulting in the existence of \rm zero amplitude points. Consequently, low light
levels and zero intensity points with phase singularities make it difficult to measure and correct the wavefront with
an AO system. \rm This is exactly the problem situation we aim to avoid with the unbalanced nulling. \rm 

\begin{figure}
\centering
\includegraphics[width=\columnwidth] {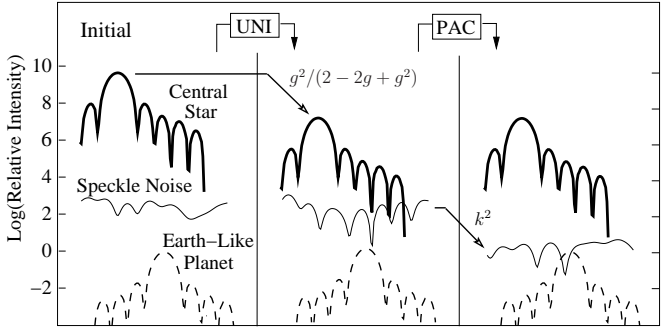}
\caption{Results of a simplified numerical simulation showing the improvement of the dynamic range in the image plane
by the UNI-PAC method. The curves show the central star (bold line) and its speckle halo noise (solid line) normalized
by the planet peak intensity (dashed line). Initially (left panel) the contrast between the star and the planet is
assumed to be $5{\times}10^{9}$ and the speckle halo lies $4{\times}10^{2}$ above the planet, which is a result of the
phase aberration of $\lambda$/1000 rms and the amplitude aberration of 0.004 rms with a flat spectrum ($30\times30$
square grids in the pupil). After the UNI is applied with the factor $g=0.1$ (middle panel), the central star intensity
is reduced by $g^2/(2-2g+g^2)$=0.0055 times. After the PAC process (right panel), the aberration-magnified wavefront is
compensated to the same aberration rms as the input. The speckle halo noise is then decreased $k^2$=0.0055 times, that
is, twice the planet peak intensity. The star profile is drawn as a point spread function of a non-modified pupil,
which can be suppressed by a downstream coronagraph.}
    \label{fig_image_profile}
\end{figure}

\subsection{Phase and amplitude correction (PAC) process }
\rm After the wavefront aberrations have been magnified by the UNI stage, the goal is to compensate
them to a level equivalent to $\varepsilon _\mathrm{A} (x,y)/E_\mathrm{A}^0$ (the initial wavefront aberration)
by a secondary AO system. \rm In order to perform the phase and amplitude correction \rm (i.e., the PAC), \rm the UNI is
followed by \rm a second \rm AO system \rm composed by two DMs (Fig.~\ref{fig_optics}) and a WFS which measures the 
wavefront shape and the intensity distribution. The compensation level can be set
differently according to the characteristics of each AO system.

The PAC process is \rm formalized by \rm applying a wavefront-aberration-reduction \rm function $h(x,y)$ \rm to the normalized 
complex aberration,
\begin{equation} \left\{ \begin{array}{lll}
     ~~~\frac{\textstyle \varepsilon _{\mathrm{PAC}}(x,y)}{\textstyle E_{\mathrm{PAC}}^0}
           &=& h(x,y)~ \frac{\textstyle  \varepsilon _{\mathrm{UNI}}(x,y) }{\textstyle E_{\mathrm{UNI}}^0} \, \\[4mm]
  \bigg\langle \bigg| \frac{\textstyle \varepsilon _{\mathrm{PAC}}(x,y)}{\textstyle E_{\mathrm{PAC}}^0} \bigg|^2 \bigg\rangle
    &=&  k^2~ \bigg\langle \bigg|\frac{\textstyle  \varepsilon _\mathrm{UNI}(x,y)}{\textstyle E _\mathrm{UNI}^0} \bigg|^2 \bigg\rangle ~,
    \label{Eq_ePAC/E0PAC}
\end{array} \right. \end{equation}
where \rm $k$ is a kind of weighted average of \rm the aberration-reduction function \rm $h(x,y)$ defined as 
\begin{eqnarray}
    k^2 &=& \frac{ \big\langle |h(x,y)~ \varepsilon _{\mathrm{UNI}}(x,y)|^2 \big\rangle}
             { \big\langle |\varepsilon _{\mathrm{UNI}}(x,y)|^2 \big\rangle} \, .
\label{Eq_k2}
\end{eqnarray}

Since the residual aberrations before and after the UNI-PAC are caused by sensing, control, \rm specifications of the 
DMs, \rm or non-common path errors of each AO system, \rm then \rm $\varepsilon _{\mathrm{PAC}}(x,y)$ or $h(x,y)$ is not correlated
with $\varepsilon _{\mathrm{UNI}}(x,y)$.

After the PAC operation, \rm the unaberrated \rm complex amplitude can \rm be 
\begin{equation} \left\{ \begin{array}{llll}
  E ^0 _{\mathrm{PAC}}  &=& E ^0 _{\mathrm{UNI}} \\[2mm] 
  I ^0 _{\mathrm{PAC}}    &=&  I ^0 _{\mathrm{UNI}}  \, .
  \label{Eq_E0PAC}
\end{array} \right. \end{equation}
In the equations above, we assumed a negligible intensity change due to the amplitude control of the PAC DMs
(under very small wavefront aberration conditions).
If we were to consider this change in intensity, we should introduce a parameter $\xi$ such that
$E ^0 _{\mathrm{PAC}}  = \xi \, E ^0 _{\mathrm{UNI}} $ with $\xi \sim 1 $ in our equations. But its effect on
the principle of the present concept should not be very large. In the following, we do not take this effect into consideration.

In the PAC output plane, \rm the complex \rm aberration term \rm becomes 
\begin{equation} \left\{ \begin{array}{lllll}
    \varepsilon _{\mathrm{PAC}}(x,y)   &=& h(x,y) ~ \varepsilon _\mathrm{UNI}(x,y)  \\[2mm]
    \big\langle I^\varepsilon_{\mathrm{PAC}}(x,y) \big\rangle &=&  k^2 \, \big\langle I^\varepsilon_\mathrm{UNI}(x,y) \big\rangle \, ,
  \label{Eq_ePAC}
\end{array} \right. \end{equation}
\rm In total \rm the complex amplitude \rm is
\begin{eqnarray}
    E _{\mathrm{PAC}}(x,y)
               = g \, E _\mathrm{A}^0 + h(x,y)~ \big(\varepsilon _\mathrm{A}(x,y) - (1-g)~ \varepsilon _\mathrm{B}(x,y) \big) \, .
    \label{Eq_EPAC}
\end{eqnarray}

For the planet, the intensity change caused by the PAC stage can be considered negligible
for reasons similar to that for Eq.(\ref{Eq_E0PAC}),
\begin{equation}
     I^{p}_\mathrm{PAC} = I^{p}_\mathrm{UNI} \,. 
    \label{Eq_IpPAC}
\end{equation}
Then the improvement of the dynamic range at the PAC stage is just the aberration component reduction expressed as, 
\begin{equation} \left\{ \begin{array}{lll}
  ~~~\frac{\textstyle ~I ^0 _\mathrm{PAC}~}{\textstyle I^{p}_\mathrm{PAC}}
      &=& \frac{\textstyle ~I^0_\mathrm{UNI}~}{\textstyle I^{p}_\mathrm{UNI}} \\[4mm]
  \frac{\textstyle \big\langle I^\varepsilon_\mathrm{PAC}(x,y)\big\rangle}{\textstyle I^{p}_\mathrm{PAC}}
      &=&  k^2 \, \frac{\textstyle \big\langle I^\varepsilon_\mathrm{UNI}(x,y)\big\rangle}{\textstyle I^{p}_\mathrm{UNI}} \, .
  \label{Eq_I0PAC/IpPAC}  
\end{array} \right. \end{equation}
If we want to express the PAC stage performance compared to the initial AO-compensated wavefront level, we can write, \rm
\begin{equation}
  \bigg\langle \bigg| \frac{\varepsilon _{\mathrm{PAC}}(x,y)}{E_{\mathrm{PAC}}^0} \bigg|^2 \bigg\rangle
    =  \eta^2~ \bigg\langle \bigg|\frac{ \varepsilon _\mathrm{A}(x,y)}{E _\mathrm{A}^0} \bigg|^2 \bigg\rangle \, ,
    \label{Eq_ePAC/E0PACeta}
\end{equation}
where ${\eta \sim 1} $ means that the \rm residual wavefront aberration after the PAC \rm is similar to that \rm delivered \rm by the
first AO system. A smaller value of $\eta$ would indicate a better compensation by the PAC \rm stage, \rm and inversely. \rm Then 
with Eqs.\,(\ref{Eq_eUNI/E0UNI}) and (\ref{Eq_ePAC/E0PAC}), we obtain 
\begin{eqnarray}
    k^2  &=& \eta^2 \, g^2/(2-2g+g^2) \, .
    \label{Eq_k2eta}
\end{eqnarray}
This expression shows that, when $\eta \sim 1$, $\, k^2$ is a number less than unity 
almost completely defined by the unbalance factor $g$ of the UNI stage. \rm

\subsection{UNI-PAC total effect}
\rm From equations (\ref{Eq_I0UNI/IpUNI}) and (\ref{Eq_I0PAC/IpPAC}), the total reduction effect on the intensities
of the unaberrated and aberrated wavefront components are respectively,
\begin{equation} \left\{ \begin{array}{lllll}
    ~~~~\frac{\textstyle \, I^0_\mathrm{PAC} \, }{\textstyle I^{p}_\mathrm{PAC}} &=&
    \frac{\textstyle g^2}{\textstyle  \, 2-2g+g^2 \, } \, \frac{\textstyle  \, I^0_\mathrm{A} \, }{\textstyle I^{p}_\mathrm{A}} \\[4mm]
    \frac{\textstyle \big\langle I^\varepsilon_\mathrm{PAC}(x,y)\big\rangle}{\textstyle I^{p}_\mathrm{PAC}}
    &=& k^2 \, \frac{\textstyle \big\langle I^\varepsilon_\mathrm{A}(x,y)\big\rangle}{\textstyle I^{p}_\mathrm{A}} \, .
   \label{Eq_IpPAC/I0PAC2A}
\end{array} \right. \end{equation}
Equivalently in the image plane, \rm the central star profile attenuation factor \rm is $g^2/(2-2g+g^2)$ \rm and the average 
speckle halo attenuation factor \rm is $k^2$. \rm These results can be seen in Fig.~\ref{fig_image_profile}, based on 
a simplified numerical simulation where the corrected wave after the PAC stage is \rm calculated \rm by generating random 
residual \rm aberrations. \rm 

The very important conclusion is that the \rm sequential combination of \rm the UNI \rm and \rm PAC \rm achieves an aberration \rm 
compensation \rm of the star wavefront amplitude \rm beyond the correction capabilities of the
individual AO systems used \rm (including the
non-common path errors). \rm We wish to call it a \textit{virtual} compensation.

With this pre-optics, what a downstream coronagraph should do is to perform the rest of the total dynamic
range, where the \rm requirements for \rm the wavefront \rm aberration \rm level and the star intensity suppression level are
relaxed by the \rm reduction \rm factors \rm of the unaberrated and the aberration wavefront (the star and the speckle) 
intensities in Eq.\,(\ref{Eq_IpPAC/I0PAC2A}). \rm 
The \rm detailed \rm contrast \rm improvement \rm in the image plane \rm by this concept \rm depends on the spectrum of the
\rm aberrations and the specifications \rm of the AO system, \rm although \rm the
(average) relative improvement of the dynamic range is the same as the pupil plane.

\begin{figure}
    \centering
    \begin{minipage}[c]{\columnwidth}
        \centering\includegraphics[width=\columnwidth]{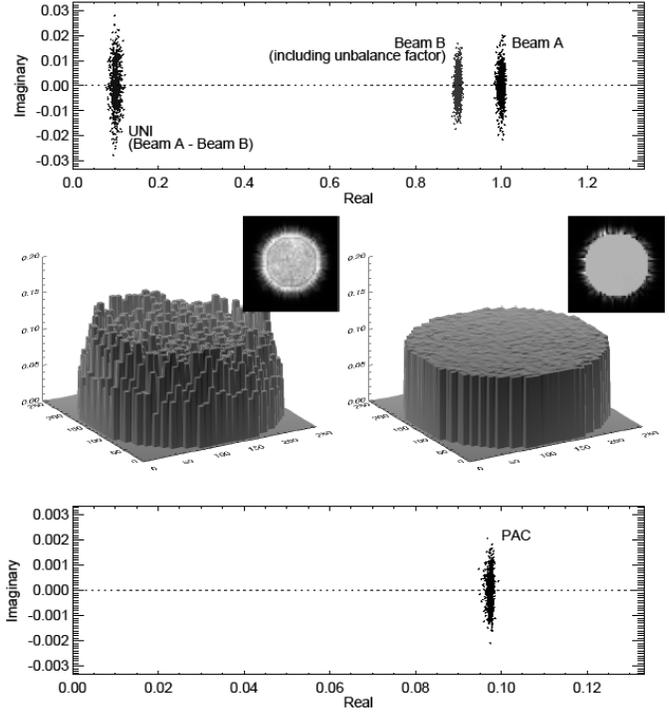}
    \end{minipage}
\caption{Simulation of the UNI-PAC resulting in a $\sim\lambda/10000$ rms from input wavefronts quality of
$\sim\lambda/1000$ rms. The top row shows the complex amplitude distribution in the complex plane for the input
Beam\,A, Beam\,B (including the unbalance factor $g=0.1$), and for the unbalanced combined beam, i.e. at the UNI output
plane. The middle row shows the amplitude (modulus) distribution of UNI output beam after the propagation between DM1
and DM2, before (left), and after (right) the Gerchberg-Saxton algorithm convergence. DM1 compensates for amplitude
fluctuations only, while DM2 compensates for the phase. The bottom row shows the PAC output wavefront complex distribution.
Note that the real and imaginary axes scale has been reduced by a factor of 1/10.}
  \label{fig:UNI_PAC_simu002}
\end{figure}

\begin{figure}
    \centering
    \begin{minipage}[c]{\columnwidth}
        \centering\includegraphics[width=\columnwidth]{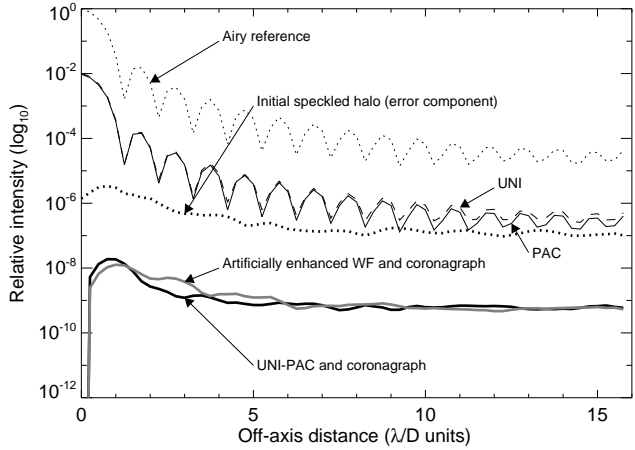}
    \end{minipage}
\caption{Azimuthally averaged profiles of focal plane images at different locations in the UNI-PAC system: a reference 
Airy profile (dotted), the profile at the UNI (dashed) and the PAC (solid) planes. The thick dotted profile 
corresponds to the speckled halo (aberration component) before the UNI stage. The two thick lines at the 
bottom of this plot are (thick black) the image profile by the \textit{virtually}-corrected PAC output wavefront with 
the ideal coronagraph, and (thick grey) the image computed by the initial wavefront with the artificial enhancement 
(i.e. with the complex aberrations reduced by a factor of $1/10$) combined with the ideal coronagraph.} 
    \label{fig:UNI_PAC_plots}
\end{figure}

\rm 

\section{Simulation of the UNI-PAC}\label{sect:simulation}
In this section we want to illustrate in a more global way how the UNI-PAC can perform, aside from purely analytical
considerations. In our simulations, we use the Fourier optics approximation, and FFT to produce complex fields at
conjugated planes. The top row of Fig.\,\ref{fig:UNI_PAC_simu002} shows the distribution of the two 
\rm complex amplitudes of the \rm beams A and B. The
initial phase \rm aberrations \rm are uniformly distributed over pupil spatial frequencies (flat power spectrum) with an rms
amplitude of $\lambda/1000$, whereas the amplitude \rm aberrations \rm follow a $f^{-2.5}$ power-law with an amplitude of
1/200$^{\mathrm{th}}$ of the average amplitude level. Beam B is shown after passing through a neutral density filter
\rm with 81\% transmission, \rm meaning that at the UNI output plane, the recombined beam
intensity is reduced by a factor $g^2=0.01$, and the phase \rm aberrations \rm are amplified.
\rm The PAC stage \rm uses two deformable mirrors (DM1 and DM2). DM1 modulates the
UNI output beam in phase only so that \rm the curvature of the wavefront induces an amplitude modification by beam 
propagation to the location of DM2. \rm Once the phase and amplitude after the UNI are measured, 
we use an FFT-based propagation algorithm combined with a recursive
Gerchberg-Saxton (\cite{gerchberg}, and e.g. \cite{fienup}) algorithm, in order to find the DM1 phase which converges
toward a given pupil amplitude distribution (flat in this case). This step is shown in the middle row of
Fig.\,\ref{fig:UNI_PAC_simu002}. The left panel shows the initial propagated beam amplitude, and the right panel shows
the same beam, but after convergence of the algorithm.
\rm DM2 compensates for the phase aberrations including those at the UNI output, as well as the effects
of phase modulation by DM1. \rm We did not use a realistic phase analysis/correction scheme. We simply
assumed that the sensing sensitivity was high enough to provide a relative phase quality similar to that used for the
input beams (Beam\,A and Beam\,B). In this simulation, \rm the DMs \rm are $32\times32$ actuators (793 within a circular
pupil are effective over a total of 1024). The distance between
the two DMs in the PAC stage was set to a relatively arbitrary distance of 30\,cm (propagation distance) while the collimated beam
diameter was 1\,cm. These parameters certainly need to be better adjusted, especially regarding the DMs dynamical
range, and the actuators influence functions, but this is not the purpose of this paper. The bottom of
Fig.\,\ref{fig:UNI_PAC_simu002} shows the final corrected wavefront at the PAC output plane. It has approximately the
same \rm wavefront aberration quality \rm as the input beams, both for phase and amplitude.

Figure\,\ref{fig:UNI_PAC_plots} shows various plots corresponding to different steps of this UNI-PAC \rm simulation. \rm The top dotted
line is the profile of the PSF computed from Beam A. For comparison purposes, the corresponding thick dotted curve is
the profile of an image given by an \rm ``ideal'' \rm coronagraph (i.e., which perfectly cancels \rm the unaberrated plane wave) \rm with
this input wavefront (around $10^{7}$ contrast level). The dashed and continuous thinner curves are respectively the
image profiles at the UNI output, and after the PAC correction. Note the small improvement of the PAC profile
(continuous line) further away in the halo \rm which means a better wavefront. \rm
Then, we applied the same \rm ``ideal'' \rm coronagraph to the PAC output,
which is shown as the thick black continuous curve at the bottom of the \rm figure, 
where the \textit{virtual} aberration correction improvement becomes obvious. \rm
For comparison, we have generated a 
similar coronagraphic image, but from the Beam\,A data of the initial wavefront, with its \rm aberrations artificially \rm attenuated by \rm a 
factor of ${1/10}$, \rm i.e. a wavefront with a phase \rm aberration \rm of $\lambda/10000$ and amplitude \rm aberrations of 1/2000. \rm 
This latter comparison demonstrates \rm a potential of the present method, \rm i.e. the \textit{virtual} \rm aberration \rm correction 
improvement \rm by the UNI-PAC is equivalent to the initial compensation. \rm

\section{Discussion}\label{sect:discussion}

\subsection{The virtual compensation}
In equation (\ref{Eq_EPAC}), at the PAC output plane, \rm both the initial complex aberrations \rm $\varepsilon
_\mathrm{A}(x,y)$ and $\varepsilon _\mathrm{B}(x,y)$ are multiplied by \rm a \rm factor $h(x,y)$. This can be interpreted as
the result of an unbalanced nulling interference of two wavefronts written as 
\begin{eqnarray} \left\{ \begin{array}{ll}
  E _\mathrm{A} (x,y) =  E _\mathrm{A}^0 + h(x,y) ~ \varepsilon _\mathrm{A} (x,y) ~ \\[2mm]
  E _\mathrm{B} (x,y) =  E _\mathrm{B}^0 + h(x,y) ~ \varepsilon _\mathrm{B} (x,y) ~.
  \label{Eq_E'AE'B}
\end{array} \right. \end{eqnarray}
In other words, it is equivalent to having a much better input wavefront ($h$ times smaller), followed by an unbalanced
nulling beam combination, but no further AO correction. \rm At \rm the UNI-PAC \rm system \rm output, one cannot
distinguish whether the wavefront was very good  at the input \rm and inside the UNI optics \rm
or if it was processed by the UNI-PAC \rm (see also Fig.\,\ref{fig:UNI_PAC_plots}). 
Thus one can consider that the present concept allows for a \textit{virtual} compensation beyond the AO performance. \rm 

\begin{figure}
    \centering
     \includegraphics[width=0.7\columnwidth]{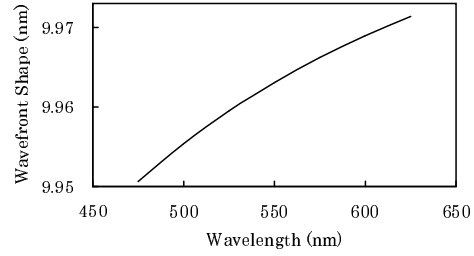}
\caption{\rm Wavelength dependence of the magnified aberration. The wavefront shape aberration after the UNI, ${z
_\mathrm{UNI}}, $ is shown for the input wavefront shape aberration ${z _\mathrm{A} = 1 \, \mathrm{nm}} $ with zero
amplitude aberration and ${g=0.1} $. In the wavelength band of 500\,nm to 600\,nm, the wavefront aberration is magnified
about 9.96 times with a small wavefront difference of about 0.01\,nm, i.e., $\lambda $/50000 level.  \rm}
    \label{fig_Achromatic}
\end{figure}

\subsection{Chromatism issues} \label{Sect_Chromatism}
In this paper the formalism is constructed for one wavelength and the method is not perfectly achromatic.

\subsubsection{Magnification process} \label{Sect_Magnification}
If the wavefront shape \rm expressed as $z(x,y)$ 
(for instance, nm unit), \rm is common \rm to every \rm wavelength \rm because of mirror reflections, 
then the phase aberration ${\theta(x,y) = 2 \, \pi \, z(x,y) \, / \lambda} $ is wavelength dependent. 

For the two beams, let us consider the wavefront shape aberrations of $z _\mathrm{A}$ 
(i.e., ${\theta _\mathrm{A} = 2 \, \pi \, z _\mathrm{A} \, /\lambda} $) and ${z _\mathrm{B} = 0} $ (this latter value is set to 
zero for convenience, but it does not affect a more general result). Wavefronts before and after the UNI are set to
\begin{eqnarray} \left\{  \begin{array}{lll}
   E _\mathrm{A}  &=&  E ^{0} + \varepsilon    \\
   E _\mathrm{B}  &=&  E ^{0}  \\
   E _\mathrm{UNI} &=& g \, E ^{0} + \varepsilon \, ,
   \label{Eq_AchromaticEA}
\end{array} \right.  \end{eqnarray}
so that the phase aberration magnification can be written (using the notation of Appendix\,\ref{Appendix_A}
and the representation of Fig.\,\ref{fig_A1})
\begin{eqnarray}
    \tan \theta_\mathrm{UNI} &=& \frac{\textstyle 1+\varepsilon _\mathrm{r} /E^{0}}
                              {\textstyle ~g+\varepsilon _\mathrm{r} /E^{0}~} \,  \tan \theta _\mathrm{A} \, ,
   \label{Eq_tantheta_UNI}
\end{eqnarray}
where $\varepsilon _\mathrm{r} $ is the real part of $\varepsilon $ and ${\theta _\mathrm{UNI}} $ is the phase
aberration after the UNI. The aberration magnification is made through the physical process in each wavelength 
and here we should check how the wavefront shape  $z _\mathrm{UNI}$ are common in the wavelength band. 
The plot in Fig.\,\ref{fig_Achromatic} shows the wavefront aberration at a point in the pupil plane 
magnified by the UNI for an initial aberration ${z _\mathrm{A} = 1 \, \mathrm{nm}} $ (and a zero amplitude
aberration), for an unbalance factor ${g=0.1} $.  The differential wavefront shape after the magnification is about 0.01\,nm
($\lambda/50000$ level), within a wavelength band between 500\,nm and 600\,nm. Here the nonlinearity
of the tangent is the main cause of the wavelength dependence. The intensity-unbalanced nulling beam combiner has \rm less dispersion,
at least for \rm an input aberration \rm level \rm of $\lambda$/1000 \rm with \rm an aberration \rm magnification factor of 10. \rm

\begin{figure}
    \centering
    \includegraphics[width=0.7\columnwidth]{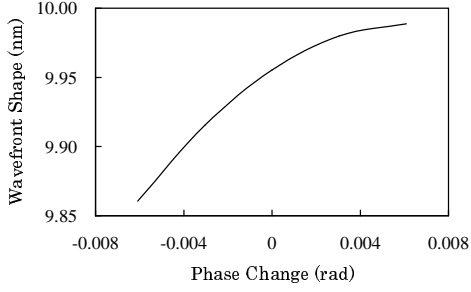}
    \caption{\rm Differential phase dependence of the magnified aberration. The wavefront shape aberration after the UNI, ${z
_\mathrm{UNI}}, $ is shown as a function of the additional phase change where the input wavefront shape aberration ${z
_\mathrm{A} = 1 \, \mathrm{nm}} $ with zero amplitude aberration under a condition of ${g=0.1}$, B = 0.2\,m, and
${\lambda=500}$\,nm. Within the phase change of $\pm\,$0.006\,rad, the wavefront aberration is magnified about 9.95 times
with a small wavefront difference of about $\pm$0.1\,nm, $\lambda $/5000 level.  \rm }
    \label{fig_Resolved}
\end{figure}

\subsubsection{Phase shifter and ND filter} \label{Sect_PhaseShifter}
Another issue of the achromaticity is for the $\pi$ phase shifter and the ND filter in the UNI stage, i.e.,
a wavelength dependence of the factor ${-(1-g)}$ (e.g., $\phi$ and $g$) in Eq.(\ref{Eq_EB''}). 

Similar to the previous section, we consider a combination of two wavefronts \rm by UNI \rm written, 
\begin{eqnarray} \left\{  \begin{array}{lll}
   E _\mathrm{A}  &=&  E ^{0} + \varepsilon    \\
   E _\mathrm{B}  &=&  \mathrm{e}^{\mathrm{i}\psi} \, E ^{0}  \\
   E _\mathrm{UNI} &=& (1-(1-g)~\mathrm{e}^{\mathrm{i}\psi}) \, E^0 + \varepsilon \, ,
   \label{Eq_ResolvedEA}
\end{array} \right.  \end{eqnarray}
wherer $\psi $ is a (wavelength-dependent) differential phase change in the wavefront B, 
then the aberration magnification is expressed as 
\begin{eqnarray}
   ~~~\frac{\textstyle \varepsilon _{\mathrm{UNI}} } {\textstyle ~ E_{\mathrm{UNI}}^0 ~ }
    &=&  \frac{\textstyle 1 }{\textstyle ~ 1-(1-g)~\mathrm{e}^{\mathrm{i}\psi} ~ } ~\frac{\textstyle  \varepsilon } {\textstyle ~ E ^0} \, .
   \label{Eq_Resolvede/E0}
\end{eqnarray}
Here the magnification factor can be calculated for a given differential phase change independent of its origin 
(such as a wavelength dependence or an incident angle to an interferometer with a baseline described in the next section).  
The magnified wavefront aberration through the UNI process as a function of the differential phase change is calculated and 
shown in Fig.\ref{fig_Resolved} for the input wavefront aberration of 1\,nm ($\lambda$/500) for phase and zero for amplitude. 
We find that the aberration is magnified to around 9.95\,nm ($\lambda$/50) with the enough small difference of about
0.1\,nm ($\lambda$/5000) within the differential phase change of $\pm$0.006\,rad ($\pm \lambda$/1000). 
The wavelength dependence of the magnification described in the previous section 
is not more critical than this term.  

Note that for a given differential amplitude change written in a range of ${g=0.1 \pm0.001}$, 
i.e., $\pm0.1\%$ difference of the transmission of the ND filter in wavelengths, 
a similar level of the differential aberration magnification can be found. 
We leave a more detailed study of these differential magnification issues for a future work. 

\rm
\subsection{Resolved stellar disk} \label{Sect_Resolved}
The formalism in Sect.~\ref{sect:formalism} treated an on-axis point source as the central star. When the source has a
resolved diameter, we should consider off-axis rays where the \rm wavefront aberrations \rm are magnified by different factors 
\rm by the UNI stage. \rm 
Considering the baseline B between the two interferometer apertures, the two wavefronts have an additional phase
difference of \rm ${\psi = 2 \pi \mathrm{B} \sin\chi \,/\,\lambda} $, where $\chi $ \rm is the incident angle of the off-axis
light. \rm 
So that we can refer the previous subsection to address this point. 

\rm For instance, with a \rm baseline of 0.2\,m (e.g., \cite{Shao04}) 0.5\,mas tilted wavefront which comes from the limb
position of the Sun at 10\,pc has a \rm phase difference of ${\psi=6.1{\times}10^{-3}}$ \rm radian \rm ($\lambda/1000$) between
the two apertures for the wavelength of 500\,nm. 
Therefore the range of the incident anlgle for the Sun at 10\,pc almost correeponds to the phase range 
shown by the curve in Fig.\ref{fig_Resolved} with the enough small differential magnification of about
$\lambda$/5000.  
This means \rm that the stellar diameter might not 
affect the performance for a short baseline interferometer \rm under small aberration conditions.
\rm 

\subsection{Larger phase aberrations}
Even for a large wavefront \rm phase aberration \rm of $\lambda$/20 and amplitude \rm aberration \rm of 0.1, the \rm present \rm method is still
effective with the factor ${g=0.3}$, where the central star extinction and speckle intensity reduction becomes about
0.05. The \rm aberration \rm level is similar to a ground-based AO condition and the gain of 0.05 is essential for exo-planet
observations from the ground, e.g., extending the observable range from
$10^{-5}$ to $5{\times}10^{-7}$. If the PAC AO reaches higher bandwidth than the telescope-side AO system (e.g.,
\cite{Guyo06}), it might be possible to apply the UNI-PAC method to ground-based observations.

\section{Conclusions}\label{sect:conclusion}
We have presented the principle of a pre-optics concept for precise wavefront \rm aberration \rm reduction in front of a
coronagraph in a terrestrial-planet finding telescope. It consists of a combination of an unbalanced nulling
interferometer and a two-deformable mirror AO system with a conventional pupil plane wavefront sensor,
\rm where it does not meet the low intensity problem for wavefront measurements. \rm This method
reduces both the source intensity as a nulling coronagraph and the (speckle halo) noise intensity produced by wavefront
\rm aberrations. \rm

\rm The wavefront magnification phenomenon at the unbalanced nulling interference makes it possible to correct
the wavefront precisely beyond the capabilities of employed AO systems with non-common path errors.
In reality, by using the UNI-PAC, the specification for all of the coronagraph optics can be relaxed
which is probably key to developing a cost-effective exo-planet detection system,
although the performance of the whole system should be investigated for each case. \rm 

\rm A candidate instrument where the UNI-PAC system can be used would be a modified Michelson interferometer
in a collimated beam of an off-axis telescope (\cite{Shao04}). \rm
We also expect the \rm present \rm concept to be applicable with several other coronagraphic concepts instead of the
UNI stage \rm by a Michelson beam combiner, \rm i.e., with more general contrast-reduction setups
(see for example \cite{Abe06}), but we leave this discussion for a future study.

Some other considerations, such as
achromaticity, resolved star nulling, planet image quality, and AO performance limitations should be investigated.

\begin{acknowledgements}
LA is supported by Grant-in-Aid (Nos. 160772048007 and 160871018002) from the Ministry of Education, Culture, Sports,
Science, and Technology (MEXT) of Japan. \rm JN and NM are supported by Grant-in-Aid 
(No. 19360036 and the Priority Areas "Development of Extra-solar Planetary Science") from the MEXT. \rm 
The authors are grateful to Profs. M.~Yoshizawa and M.~Tamura for their constant
encouragements with this work.
\end{acknowledgements}

\rm
\begin{appendix}
\section{Wavefront aberration expressions\label{Appendix_A}}

\begin{figure}
    \centering
    \includegraphics[width=0.6\columnwidth]{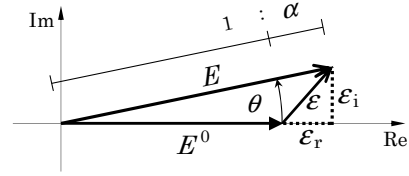}
    \caption{\rm Correspondence of the electric field definitions.  \rm }
    \label{fig_A1}
\end{figure}

The complex amplitude of an electric field in a pupil plane with a ``classical'' wavefront phase aberration $\theta (x,y) $
and an amplitude aberration $\alpha (x,y) $ can be written as 
\begin{eqnarray} \label{AEq_E}
  E(x,y) &=& E ^{0}~ \bigl( 1+\alpha (x,y) \bigr) \ \mathrm{e}^{\mathrm{i}\theta (x,y)} \, .
\end{eqnarray}
Here $E^{0} $ is the unaberrated amplitude (usually assumed as a real number), and the effect of the aberrations can be
defined as a deviation $\varepsilon$ between $E$ and $E^{0}$ noted,
\begin{eqnarray} \label{AEq_e}
  \varepsilon (x,y) = E(x,y) - E ^{0} \, ,
\end{eqnarray}
which is the complex aberration component in Eq.(\ref{Eq_E}), at the beginning of the formalism in Sect\,\ref{sect:formalism}.
Relationships between $\varepsilon $ and the ``classical'' wavefront aberrations can be seen in Fig.\ref{fig_A1}, written as
\begin{eqnarray} \left\{  \begin{array}{lll}
    \tan \theta(x,y) &=& \frac{\textstyle \varepsilon _\mathrm{i}/E^{0}}
                              {\textstyle ~ 1+ \varepsilon _\mathrm{r} /E^{0} ~}  ~ \\[3mm]
    \alpha(x,y)  
               &=&  \sqrt{ (1+\varepsilon _\mathrm{r} /E^{0} )^2 +  (\varepsilon _\mathrm{i} /E^{0} )^2 } \, - 1 \, ,
  \label{Eq_tantheta}
\end{array} \right.  \end{eqnarray}
and
\begin{eqnarray} \left\{  \begin{array}{lll}
     \varepsilon _\mathrm{i}/E^{0} &=& \big( 1+\alpha(x,y) \big) \, \sin \theta(x,y) ~ \\[2mm]
     \varepsilon _\mathrm{r}/E^{0} &=& \big( 1+\alpha(x,y) \big) \, \cos \theta(x,y) \, - \, 1 \, \\[2mm]
    \varepsilon (x,y) &=& \varepsilon _\mathrm{r} + \mathrm{i} \, \varepsilon _\mathrm{i}
  \label{Eq_ei/E0}
\end{array} \right.  \end{eqnarray}
where $\rm{i} $ is the imaginary unit. $\varepsilon _\mathrm{r} $ and $\varepsilon _\mathrm{i} $ are real numbers which 
indicate the real and the imaginary part of $\varepsilon (x,y) $, respectively, and where $(x,y)$ is not indicated for simplicity.

$E^{0} $ can be defined as a spatial ensemble average of
$E$ within the pupil plane, ${E^{0} = \langle E(x,y) \rangle} $ so that the aberration component has a zero mean, ${\langle
\varepsilon (x,y) \rangle = 0} $, whereas ${\langle \alpha (x,y) \rangle} $ and ${\langle \theta (x,y) \rangle} $
are not always necessarily zero.
Under very small wavefront aberration conditions ${|\varepsilon/E^{0}| \ll 1} $, as for the present method, approximations
\begin{eqnarray} \left\{  \begin{array}{lll}
    \theta(x,y) & \sim & \varepsilon _\mathrm{i} / E^{0}  ~ \\[1mm]
    \alpha(x,y)  & \sim & \varepsilon _\mathrm{r} / E^{0}
  \label{Eq_theta_alpha}
\end{array} \right.  \end{eqnarray}
can be used. Therefore the normalized complex aberration, $\varepsilon (x,y)/E^{0}$ is a good indicator
of the wavefront aberrations for our concept.

\section{Correlation between the two wavefront aberrations at the UNI stage\label{Appendix_B}}
At the UNI stage, if the complex aberrations of the two wavefronts strongly correlate
and have almost the same amplitude, $\varepsilon _\mathrm{A}(x,y) \sim \varepsilon _\mathrm{B}(x,y)$, the
difference between the two wavefronts at planes A and B tends to zero.
Then the aberration component of the UNI output becomes
\begin{eqnarray} \left\{  \begin{array}{lll} 
    \varepsilon _\mathrm{UNI}(x,y) &\sim& g \, \varepsilon _\mathrm{A}(x,y)  ~ \\[1mm]
    \big\langle I^\varepsilon_{\mathrm{UNI}}(x,y) \big\rangle
                                               &\sim&   g^2 \, \big\langle I^\varepsilon_\mathrm{A}(x,y) \big\rangle \, ,
  \label{Eq_correlated} 
\end{array} \right.  \end{eqnarray}
instead of Eq.\,(\ref{Eq_eUNI}). This means that a simple nulling interference can reduce both the star intensity
and the speckled halo. This constitutes a better situation for achieving a deeper nulling without taking being 
concerned with the aberration, but it would be unrealistic. 

If both wavefront aberrations are perfectly uncorrelated, the unbalanced nulling combination produces
a larger complex aberration term. This is the "worst case" situation expressed by Eq.\,(\ref{Eq_eUNI}).

When the two wavefronts are taken at different positions in a telescope primary or a telescope-side AO
system, they may partially correlate and the nulling combination will provide results intermediate between those
described above.

\section{Planet intensity after the UNI stage\label{Appendix_C}}
We examine how an off-axis object (a planet) is affected by the UNI stage. 
We are working in an interferometric beam combination scheme (two sub-pupils
extracted from a monolithic telescope mirror, or using diluted apertures). During a ``blind search'' observation, we
must consider the rotation of the nulling interferometer to explore the planet in every direction, which results in an
averaging of the planet signal between the (pseudo-)nulled and constructive conditions of the interferometer for a
given baseline $B$ (e.g., \rm \cite{Brac78}). \rm Here the planet should also be sufficiently off-axis, i.e. more than 
$\sim \lambda/B$ to prevent the planet staying in the central nulled region. 

\rm We introduce the complex amplitude of the planet \rm ${E^{p}_\mathrm{A} \, (=\!E^{p}_\mathrm{B})}$ \rm and its
intensity \rm $I^p_A $, where we only consider the unaberrated amplitude, and neglect the aberration term, since its
intensity is very small. \rm The average of the \rm planet intensities for the \rm constructive, \rm 
${| \, E^{p}_\mathrm{A} \, -\,(1-g) \, E ^{p} _\mathrm{B} \, |^2} $, and the destructive,
${| \, E ^{p} _\mathrm{A} \, + \, (1-g) \, E ^{p}_\mathrm{B} \, |^2} $, \rm conditions \rm after the UNI becomes
\begin{eqnarray}
     I^{p}_\mathrm{UNI} = (2-2g+g^2)~ I ^{p} _\mathrm{A} \, ,
  \label{A_IpUNI}
\end{eqnarray}
which should be used for calculation of the dynamic range improvement.

\end{appendix}
%
\rm 

\vspace{10mm}

%
%
\end{document}